\documentclass[aps,nofootinbib,longbibliography,prd,twocolumn]{revtex4-1}
\usepackage[babel]{csquotes}
\usepackage{graphicx}
\usepackage{amsmath,amssymb}
\usepackage[colorlinks,citecolor=blue,linkcolor=blue,urlcolor=blue]{hyperref}
\usepackage{mathrsfs}
\usepackage{enumerate}
\def\be{\begin{equation}}
\def\ee{\end{equation}}

\def\bra#1{\mathinner{\langle{#1}|}}
\def\ket#1{\mathinner{|{#1}\rangle}}

\usepackage{bbm}
\usepackage{caption}

\DeclareMathOperator{\Tr}{Tr}

\begin{document}

\title{On the possibility of experimental detection of the discreteness of time}

\author{Marios Christodoulou and Carlo Rovelli}

\affiliation{\small
 \mbox{}\\ 
\mbox{}
}
\date{\small\today}

\begin{abstract}
\noindent The Bose-Marletto-Vedral experiment tests a non-relativistic quantum effect due to a gravitational interaction. It has received attention because it may soon be within observational reach in the lab. We point out that: (i)  in relativistic language the experiment tests an interference effect between proper-time intervals; (ii)  the relevant difference of proper times approaches the Planck time if the masses of the particles in the experiment approach the Planck mass ($\sim$micrograms). Therefore the experiment might open a window on the structure of time at the Planck scale. If time differences are discrete at this scale ---as quantum gravity research may suggest--- the Planckian discreteness of time could show up as quantum levels of a measurable entanglement entropy. 

\noindent 

\end{abstract}

\maketitle

\section{Introduction}

Bose {\em et.\,al.}\,\cite{Bose2017a} and Marletto and Vedral \cite{Marletto2017a,Marletto2017} have  proposed an ingenious idea to amplify and observe minuscule non-relativistic quantum gravitational effects in a table-top experiment.  The idea has received considerable attention \cite{Anastopoulos2018,Hall2018,Belenchia2018,Giampaolo2018,Marletto:2018lsb,Albers2008,Christodoulou2018a,Reginatto2018,Carney2018a,Belenchia2018a,Balushi2018}. In the version proposed in \cite{Bose2017a} the observable signal is given by Bell--like correlations among the spins of two particles. The correlations are produced by a gravitational interaction. Taking into account the identification of gravity with spacetime geometry  assumed by general relativity, the observation of these correlations implies then that spacetime geometry can be in a quantum superposition (in a non-semiclassical state), and therefore can be taken as evidence for quantum behaviour of the geometry \cite{Christodoulou2018a}. 

The Bose-Marletto-Vedral (BMV) effect is predicted by low energy perturbative quantum gravity, and therefore by any approach to quantum gravity consistent with this low energy expansion, including string theory and loop quantum gravity.  It is therefore plausibly real. If detected, it would provide indirect empirical evidence that spacetime geometry does obey quantum mechanics.

On the other hand, the BMV effect is insensitive to the limit $c\to \infty$, hence it is non-relativistic. For this reason, it does not test the full relativistic quantum gravitational regime. In fact, it can be accounted for purely in terms of the scalar non-radiative modes of the gravitational field, hence it does not test the quantum \emph{dynamics} of gravity. If we do not fold the relativistic information provided by classical general relativity in, we can interpret the effect in terms of action at a distance, and discount its relevance for the quantum properties of spacetime, as emphasized for instance in \cite{Anastopoulos2018}. 

However, we point out here that a refinement of the BMV effect \emph{could} open a true window on a genuine relativistic quantum gravitational affect: time discreteness. This would definitely not be accounted for by non-relativistic quantum physics. 

The reason this is possible is that (in its relativistic interpretation) the BMV set up is a delicate interference apparatus that picks up a tiny difference $\delta \tau$ in proper time between two quantum branches. If the experiment is performed with particles of mass $m$, the phase difference $ \delta\phi$ it measures can be written as (see equation (16) in reference \cite{Christodoulou2018a})
\be
         \delta\phi=\frac{m}{m_{Pl}}\ \frac{\delta \tau}{t_{Pl}},
         \label{uno}
\ee
where $m_{Pl}$ and $t_{Pl}$ are the Planck mass and the Planck time.  The Planck mass is an easily accessible scale ($\sim$micrograms), while $t_{Pl}$ is at the so-far deeply inaccessible scale $t_{Pl}\!\sim\!10^{-44}s$.  But an interference apparatus detecting $\delta\phi$ of order unit using particles with masses getting closer to $m_{Pl}$ tests $\delta \tau$ at scales approaching the Planck time $t_{Pl}$.   

Now, it is often pointed out in quantum gravity research that the Planck time $t_{Pl}$ could be a minimal observable time; this follows for instance from relativity plus the fact that many approaches to quantum gravity predict a minimal length \cite{Garay1994}.  The simplest possibility is to assume that a measurement of a time lapse can only yield multiples of the Planck time. If this holds for the time difference $\delta t$, namely if $\delta \tau=n\; t_{Pl}$, with integer $n$, then 
\be
         \delta\phi=n \frac{m}{m_{Pl}}. 
\ee
As discussed below, such a discretisation of the phase could be observable if $m$ approaches the scale of $m_{Pl}$.  The current proposal to measure the BMV effect in the lab relies on the use of nano-particles, with masses of order $10^{-6}m_{Pl}$. With these particles the BMV effect tests $\delta\tau$ of the order of $10^6 {t_{Pl}}$. This is too large to hope to see time discreteness. But if the experiment can be pushed to work with larger particles, with masses getting closer to $m_{Pl}$, then a discontinuity in $\delta\tau$ would reflect in a discontinuity in $\delta\phi$.  In turn, the interference phase $\delta\phi$ can be detected by the Bell--like correlations among the particles' spins, and these can acquire a characteristic quantum band structure.

This analysis is rough and the effect might be questioned by a more detailed investigation. For instance, it may turn out that the BMV apparatus does not measure eigenvalues but rather expectation values, or that the scale of discreteness for differences in duration is actually smaller that Planckian.  Still, a  prospect of experimental access to the scale of the Planck time is so interesting to deserve full attention. 

\section{The BMV experiment}

Let us start by describing the BMV experiment in relativistic language, as in \cite{Christodoulou2018a}. Two particles ($a$ and $b$) of mass $m$ and spin $\frac12$ are quantum split (say with a Stern-Gerlach-like apparatus) and each is set in a superposition of two distinct states, say with spins $+$ and $-$ in some basis, with different positions in space.  This gives rise to four different branches, which we denote $|++\rangle,|+-\rangle,|-+\rangle$ and $|--\rangle$ and a tensor state 
\begin{eqnarray}
\ket{\psi} &=&
\frac{\ket{+}_a+\ket{-}_a}{\sqrt2} \otimes \frac{\ket{+}_b+\ket{-}_b}{\sqrt2}
\nonumber \\&= &
\frac{ \ket{++}+\ket{+-}+\ket{-+}
+\ket{--}}2.
\end{eqnarray}

After a time $t$ the two components of each particle are recombined. The relative positions of the particles differ in the distinct branches during the time $t$, giving rise to different gravitational fields, namely different spacetime geometries. Therefore during the interval $t$ the quantum state of the geometry is in a superposition of four (semiclassical) spacetimes, each corresponding to a classical metric. In particular, the proper time $\tau$ along the worldline of one particle is affected by the presence of the other by relativistic time dilation.  This effect is obviously very small, but, as we shall see, it may be picked up by interference. 

\begin{figure}[t]
\includegraphics[scale=0.4]{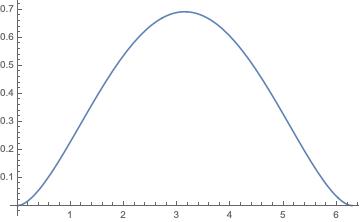}
\caption{The entanglement entropy for $\delta \phi \in \{0, 2 \pi\}$.}
\label{fig:entropy}
\end{figure}

For simplicity, consider the case in which the two particles are kept at a small distance $d$ only in a single branch, say $|--\rangle$, while in the other three branches the time dilation is negligible.  According to general relativity, the gravitational time dilation is  
\be
\delta \tau=\frac{Gm}{dc^2}\; t. \label{a}
\ee
where $G$ is the Newton constant and $c$ the speed of light.  The phase of the quantum state of a particle of mass $m$ evolves in time as $e^{i\phi}=e^{im c^2\tau/\hbar}$. Therefore after a time $t$ the $|--\rangle$ branch picks up a phase difference 
\be
\delta\phi=\frac{m c^2}{\hbar}\delta\tau \label{b}
\ee
with respect to the other branches. This equation is equivalent to equation \eqref{uno}.  After the time $t$ the state of the two particles has become
\begin{equation}
\ket{\psi} =\frac{\ket{++}+\ket{+-}+\ket{-+}
+e^{i \delta \phi}\ket{--}}2
.\end{equation}
This is an entangled state.  The amount of entanglement is measured by the entanglement entropy 
\be
I=\Tr[\rho\ln\rho]
\ee
where
\begin{equation}
\rho= \Tr_b \ket{\psi}\bra{\psi}
\end{equation}
the trace being on the spin states of one of the two particles.  A quick calculation gives
\begin{align}
\rho&= \frac{1}{2} \bigg( \ket{+}\bra{+} + \ket{-}\bra{-} \bigg) \nonumber \\
&+\frac{e^{-i \delta \phi}+1}{4} \ket{+}\bra{-}+\frac{e^{i \delta \phi}+1}{4}\ket{-}\bra{+}.
\end{align}
This is correctly a hermitian matrix of unit trace. To compute the entropy we need to diagonalise $\rho$. A straightforward calculation gives the eigenvalues 
\be
\rho_\pm = \frac12 \pm \frac{\sqrt{1+\cos \delta \phi}}{2\sqrt{2}}
\ee
When $\delta \phi =0$, $\rho_+=1$ and $\rho_-=0$, thus giving vanishing entanglement entropy, i.e. there is no interference in the output. When $\delta \phi =\pi$, $\rho_+=1/2$ and $\rho_-=1/2$; the state is maximally entangled and $I= \log2$, i.e. we observe the BMV effect.  For a general $\delta\phi$, the entanglement entropy is 
\begin{eqnarray}
I&=&-\rho_+\ln \rho_+ - \rho_-\ln \rho_-   \\ \nonumber
&=&- \left(\frac12 + \frac{\sqrt{1+\cos \delta \phi}}{2\sqrt{2}}\right)\ln\left(\frac12 + \frac{\sqrt{1+\cos \delta \phi}}{2\sqrt{2}}\right)
\\ \nonumber && -
\left(\frac12 - \frac{\sqrt{1+\cos \delta \phi}}{2\sqrt{2}}\right)\ln \left(\frac12 - \frac{\sqrt{1+\cos \delta \phi}}{2\sqrt{2}}\right).
\end{eqnarray}
See Figure \ref{fig:entropy}.  In the lab, $\delta\phi$ can be controlled by modulating $t$, via
\be
\delta\phi=\frac{Gm^2}{d\hbar}\; t. 
\ee
that follows from \eqref{a} and \eqref{b}. The entanglement entropy can be measured by the violation of the Bell inequalities in repeated spin measurements on the recombined particles.  

\begin{figure}[t]
\includegraphics[scale=0.4]{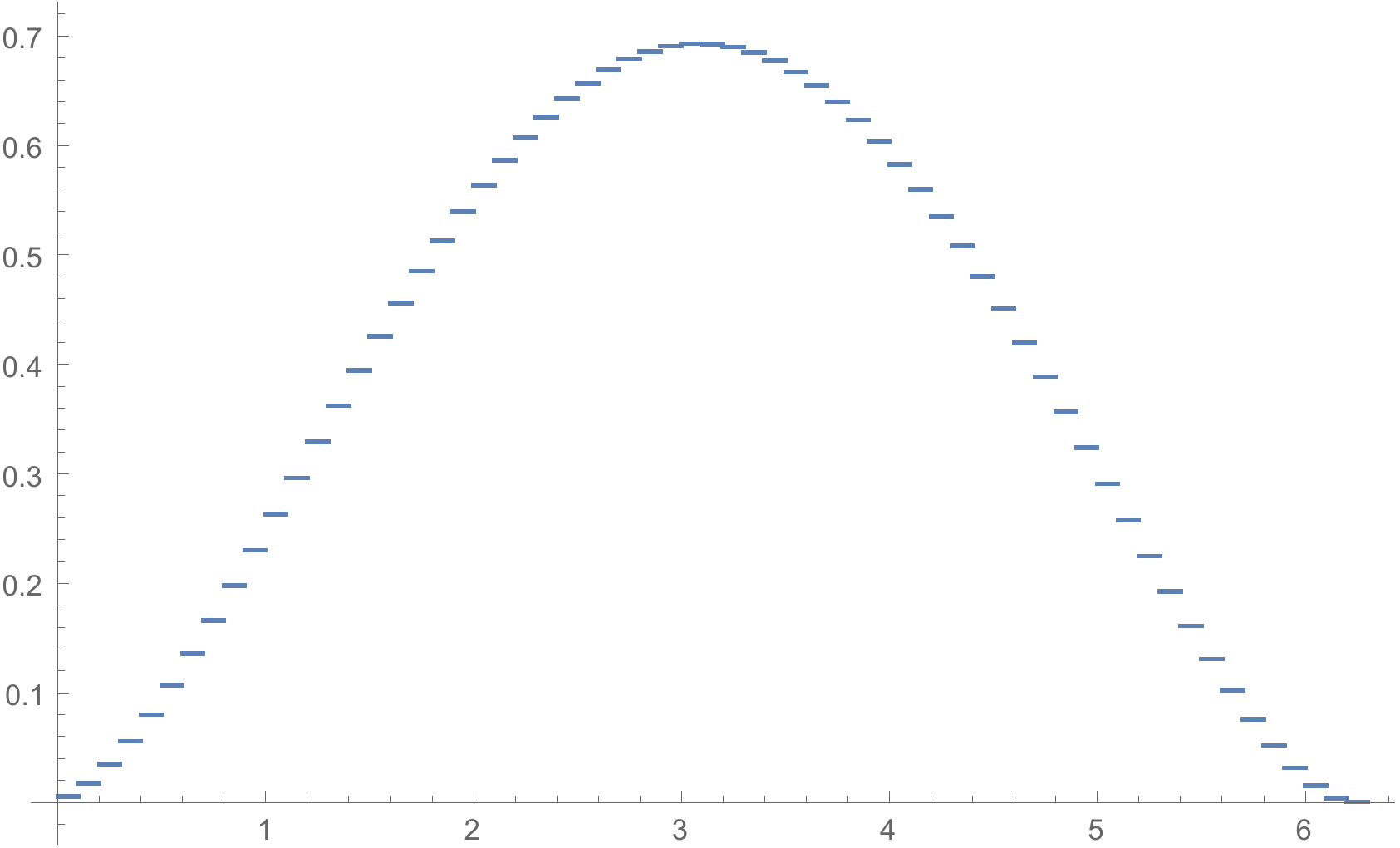}
\caption{The entanglement entropy for $\delta \phi \in \{0, 2 \pi\}$ under the assumption that $\delta t/t_{Pl} \in \mathbb{N}^+$, for particles with $1/10$ of the Planck mass.}
\label{fig:discrEntropy}
\end{figure}

Consider now the hypothesis that time is discrete at the Planck scale. We consider here for simplicity the simplest possible ansatz: that 
\begin{equation}
 \delta \tau \in \mathbb{N}
\end{equation}
 in natural units. That  is, $\delta \tau=n\, t_P$ with a non negative integer $n$.  Writing $m=\alpha \, m_P$ with $\alpha$ a dimensionless real parameter, we have that the only values of $\phi$ that are actually realized are 
\begin{equation} \label{eq:discretePhi}
\delta \phi = \alpha \, n,
\end{equation}
that is, the phase ends up taking only discrete quantised value, when $t$ is varied continuously.  It follows that the entropy is not anymore given by a continuous curve as in Figure \ref{fig:entropy}, but has characteristic quantum steps. As long as $\alpha\ll 1$, as in the realization of the BMV experiment currently in preparation, the steps are too fine to be resolved, but if $\alpha$ approaches unit the steps become visible, as as in Figure \ref{fig:discrEntropy}, where $\alpha=.1$. 

For particles with masses larger that the Planck mass interference is likely to disappear altogether, as is common in interference experiments when the wave frequency is much higher than the relevant scale of the apparatus. In this case wave theory goes to the eikonal approximation. Wave mechanics goes to classical mechanics. The Compton frequency 
\be
\nu_c  = \frac{m c^2}{\hbar}=    \frac{m}{m_{Pl}}\ \frac{\nu_P}{2\pi } 
\ee
of objects with mass larger than the Planck mass is formally larger than the Planck frequency $\nu_{Pl}=\frac{2\pi}{t_{Pl}}$ and probably meaningless.  
 
 Notice that in this case an apparatus capable of detecting $\delta\phi\sim 1$ is going to be affected by genuine dynamical effects since we can also write
\be
\delta\phi  = \frac{m^2}{m^2_{Pl}} \frac{ct}{d}
\ee
and if the left hand side and the first fraction are of order unit, so must be the second, with the consequence that the duration $t$ of the interaction must be of the same order than the light travel time $d/c$ between the particles (see also \cite{Mari2016}). This would take us outside the static approximation used in the analysis.\\[4mm]

\section{Discussion}

The current hope is to realise the BMV experiment in the lab with masses $m \sim 10^{-6} m_P$ in the next few years \cite{Bose2017a}.  Already with masses at this scale, the BMV experiment is testing time differences of the order of  $\delta \tau  \sim 10^{-38}s \sim 10^{6} t_{Pl}$. This is already an extraordinarily small time.  For comparison, the most accurate direct measurements of time at our disposal make use of the frequencies corresponding to energy differences in atomic states, atomic clocks, with an accuracy corresponding at best to a period of the order $\sim 10^{-19}s$ \cite{Marti2018}, namely about twenty orders of magnitude larger. 

A relativistic language is not needed to derive the correlations that the BMV experiment is expected to detect. In the non-relativistic language no small time intervals are in play: instead of $\delta\phi=mc^2\delta \tau/\hbar$, the phase reads $\delta\phi=t\, \delta E/\hbar$, and the $c^2$ makes all the difference.  But if time discreteness is detected, the non-relativistic language becomes insufficient to descrive the relevant physics: time discreteness is a genuine relativistic quantum gravitational effect.   

As mentioned in the introduction, the analysis given here assumes the discreteness of $\delta\tau$ in Planck time multiples. It is possible, but it is not certain that this is implied by quantum gravity.  Two reasons that could question this assumption are the following. First, the spectrum of $\tau$ could be less trivial and, as a consequence, \emph{differences} of proper time could be much smaller.  For instance, if the spacing between eigenvalues decreases  when the eigenvalues are large, their differences may become small.  Second, a more careful analysis might show that the interference depends on averages, or expectation values of time durations, and these may be continuous even if direct duration measurements are quantised.  

Even with these caveats, the possibility that quantum interference effects could depend on time differences of the order of Planck time, a scale so far considered totally out of reach, definitely deserves attention.  

\appendix

\section*{Acknowledgments}
The authors acknowledge support from the kind donors to the Samy Maroun Center for Space, Time and the Quantum.

\vfill

\bibliographystyle{utcaps}
\bibliography{/Users/carlorovelli/Documents/library}

\providecommand{\href}[2]{#2}\begingroup\raggedright\begin{thebibliography}{10}

\bibitem{Bose2017a}
S.~Bose, A.~Mazumdar, G.~W. Morley, H.~Ulbricht, M.~Toro{\v{s}},
  M.~Paternostro, A.~A. Geraci, P.~F. Barker, M.~S. Kim, and G.~Milburn,
  ``{Spin Entanglement Witness for Quantum Gravity},'' {\em Physical Review
  Letters} {\bf 119} (2017) no.~24, 240401,
  \href{http://arxiv.org/abs/1707.06050}{{\tt arXiv:1707.06050}}.

\bibitem{Marletto2017a}
C.~Marletto and V.~Vedral, ``{Witness gravity's quantum side in the lab},''
  \href{http://dx.doi.org/10.1038/547156a}{{\em Nature} {\bf 547} (2017)
  no.~7662, 156--158}.

\bibitem{Marletto2017}
C.~Marletto and V.~Vedral, ``{Gravitationally Induced Entanglement between Two
  Massive Particles is Sufficient Evidence of Quantum Effects in Gravity},''
  {\em Physical Review Letters} {\bf 119} (2017) no.~24, 240402,
  \href{http://arxiv.org/abs/1804.11315}{{\tt arXiv:1804.11315}}.

\bibitem{Anastopoulos2018}
C.~Anastopoulos and B.-L. Hu, ``{Comment on ``A Spin Entanglement Witness for
  Quantum Gravity"; and on ``Gravitationally Induced Entanglement between Two
  Massive Particles is Sufficient Evidence of Quantum Effects in Gravity"},''
  \href{http://arxiv.org/abs/1804.11315}{{\tt arXiv:1804.11315}}.

\bibitem{Hall2018}
M.~J. Hall and M.~Reginatto, ``{On two recent proposals for witnessing
  nonclassical gravity},'' {\em Journal of Physics A: Mathematical and
  Theoretical} {\bf 51} (2018) no.~8, 085303,
  \href{http://arxiv.org/abs/1707.07974}{{\tt arXiv:1707.07974}}.

\bibitem{Belenchia2018}
A.~Belenchia, R.~Wald, F.~Giacomini, E.~Castro-Ruiz, {\v{C}}.~Brukner, and
  M.~Aspelmeyer, ``{Quantum Superposition of Massive Objects and the
  Quantization of Gravity},'' {\em ArXiv: 1807.07015} (2018)  ,
  \href{http://arxiv.org/abs/1807.07015}{{\tt arXiv:1807.07015}}.

\bibitem{Giampaolo2018}
S.~M. Giampaolo and T.~Macr{\`{i}}, ``{Entanglement, holonomic constraints, and
  the quantization of fundamental interactions},''
  \href{http://arxiv.org/abs/1806.08383}{{\tt arXiv:1806.08383}}.

\bibitem{Marletto:2018lsb}
C.~Marletto and V.~Vedral, ``{When can gravity path-entangle two spatially
  superposed masses?},'' \href{http://arxiv.org/abs/1803.09124}{{\tt
  arXiv:1803.09124}}.

\bibitem{Albers2008}
M.~Albers, C.~Kiefer, and M.~Reginatto, ``{Measurement analysis and quantum
  gravity},'' {\em Physical Review D} {\bf 78} (2008) no.~6, 064051,
  \href{http://arxiv.org/abs/0802.1978}{{\tt arXiv:0802.1978}}.

\bibitem{Christodoulou2018a}
M.~Christodoulou and C.~Rovelli, ``{On the possibility of laboratory evidence
  for quantum superposition of geometries},''
  \href{http://arxiv.org/abs/1808.05842}{{\tt arXiv:1808.05842}}.

\bibitem{Reginatto2018}
M.~Reginatto and M.~J.~W. Hall, ``{Entanglement of quantum fields via classical
  gravity},'' \href{http://arxiv.org/abs/1809.04989}{{\tt arXiv:1809.04989}}.

\bibitem{Carney2018a}
D.~Carney, P.~C.~E. Stamp, and J.~M. Taylor, ``{Tabletop experiments for
  quantum gravity: a user's manual},''
  \href{http://arxiv.org/abs/1807.11494}{{\tt arXiv:1807.11494}}.

\bibitem{Belenchia2018a}
N.~Kosterina, R.~Wang, A.~Eriksson, and E.~M. Gutierrez-Farewik, ``{Force
  enhancement and force depression in a modified muscle model used for muscle
  activation prediction},'' {\em Journal of Electromyography and Kinesiology}
  {\bf 23} (2013) no.~4, 759--765, \href{http://arxiv.org/abs/1807.07015}{{\tt
  arXiv:1807.07015}}.

\bibitem{Balushi2018}
A.~AlBalushi, W.~Cong, and R.~B. Mann, ``{Optomechanical quantum Cavendish
  experiment},'' {\em Physical Review A} {\bf 98} (2018) no.~4, 043811,
  \href{http://arxiv.org/abs/1806.06008}{{\tt arXiv:1806.06008}}.

\bibitem{Garay1994}
L.~J. Garay, ``{Quantum gravity and minimum length},'' {\em Int.J.Mod.Phys.}
  {\bf A10} (1994)  145--166, \href{http://arxiv.org/abs/9403008}{{\tt
  arXiv:9403008 [gr-qc]}}.

\bibitem{Mari2016}
A.~Mari, G.~{De Palma}, and V.~Giovannetti, ``{Experiments testing macroscopic
  quantum superpositions must be slow},''{\em Scientific Reports} {\bf 6} (sep,
  2016)  22777, \href{http://arxiv.org/abs/1509.02408}{{\tt arXiv:1509.02408}}.

\bibitem{Marti2018}
G.~E. Marti, R.~B. Hutson, A.~Goban, S.~L. Campbell, N.~Poli, and J.~Ye,
  ``{Imaging Optical Frequencies with 100 $\mu$hz Precision and 1.1 $\mu$ m
  Resolution},'' {\em Physical Review Letters} {\bf 120} (2018)  ,
  \href{http://arxiv.org/abs/1711.08540}{{\tt arXiv:1711.08540}}.

\end{thebibliography}\endgroup

\end{document}